\documentclass[
  journal=pasa,
  manuscript=research-paper, 
  year=xxxx,
  volume=xx,
]{cup-journal}

\usepackage{microtype,booktabs,amsmath}

\usepackage[breaklinks,colorlinks,
   urlcolor=blue,citecolor=blue,linkcolor=blue]{hyperref}

\newcommand{\fig}{Fig.}

\newcommand{\tab}{Table}

\newcommand{\sect}{Section}

\newcommand{\eqn}{Equation}
\newcommand{\eqns}{Equations}

\newcommand{\Fig}{Fig.}

\def\degr     {\hbox{$^\circ$}}
\def\arcmin   {\hbox{$^{\prime}$}}

\title{A measurement of small-scale features using ionospheric scintillation. Comparison with refractive shift measurements}

\author{A.~Waszewski}
\affiliation{International Centre for Radio Astronomy Research, Curtin University, Bentley, WA6102, Australia}
\email[A. Waszewski]{angelica.waszewski@postgrad.curtin.edu.au}
\author{J.~Morgan}
\affiliation{International Centre for Radio Astronomy Research, Curtin University, Bentley, WA6102, Australia}
\author{C.~H.~Jordan}
\affiliation{International Centre for Radio Astronomy Research, Curtin University, Bentley, WA6102, Australia}
\alsoaffiliation{ARC Centre of Excellence for All Sky Astrophysics in 3 Dimensions (ASTRO 3D), Bentley, Australia}

\doi{10.1017/pas.2022.xxx.}

\received {dd Mmm YYYY}
\revised  {dd Mmm YYYY}
\accepted {dd Mmm YYYY}
\published{dd Mmm YYYY}


\begin{document}
\begin{abstract}

We present a study of scintillation induced by the mid-latitude ionosphere. By implementing methods currently used in Interplanetary Scintillation studies to measure amplitude scintillation at low frequencies, we have proven it is possible to use the Murchison Widefield Array to study ionospheric scintillation in the weak regime, which is sensitive to structures on scales $\sim$\,300 m at our observing frequency of 154 MHz, where the phase variance on this scale was 0.06 rad\textsuperscript{2} in the most extreme case observed. Analysing over 1000 individual 2-minute observations, we compared the ionospheric phase variance with that inferred with previous measurements of refractive shifts, which are most sensitive to scales almost an order of magnitude larger. The two measurements were found to be highly correlated (Pearson correlation coefficient 0.71). We observed that for an active ionosphere, the relationship between these two metrics is in line with what would be expected if the ionosphere's structure is described by Kolmogorov turbulence between the relevant scales of 300\,m and 2\,000\,m. In the most extreme ionospheric conditions, the refractive shifts were sometimes found to underestimate the small-scale variance by a factor of four or more, and it is these ionospheric conditions that could have significant effects on radio astronomy observations. 

\end{abstract}

\section{INTRODUCTION}
In recent years there has been a resurgence in low-frequency radio astronomy.
LOFAR \citep{vanhaarlem2013} and the Murchison Widefield Array \citep{tingay2013mwa} in particular have proven to be excellent tools for probing of the ionosphere.
In a series of papers, \citet{cleo2015b, cleo2015c, cleo2015a, cleo2016a, cleo2016b} demonstrated the capability of the Murchison Widefield Array to measure exquisitely the 2D spatial derivatives of the Total Electron Content (TEC) of the ionosphere via measurements of the refractive shifts of the several hundred sources visible in a typical MWA snapshot image.
A wide range of phenomena were observed, including Travelling Ionospheric Disturbances (TIDs) and magnetic field aligned structures which were shown (via parallax imaging) to be located in the magnetosphere.
This method of refractive shift measurement, and statistics derived from it has become the de facto metric by which ionospheric activity is measured in MWA observations \citep{jordan2017}.
They have been studied extensively \citep{2020RaSc...5507106H}, and methods have also been developed to mitigate the effects of refractive distortion for imaging \citep{2018A&C....25...94H} and source astrometry \citep{morgan2018}.

Meanwhile, \citet{mevius2016} have used LOFAR to measure the ionosphere using phase calibration solutions.
This results in a measurement of phase variance per baseline length, allowing the structure function (see \sect~\ref{sec:turbulence}) of the ionosphere to be constructed for baselines from 0.1--100\,km.
A similar approach has been used by \citet{2022JATIS...8a1012R}, to use the long-baseline configuration of Phase-II MWA \citep{2018PASA...35...33W,2019PASA...36...50B} to probe the ionosphere down to scales of around 600\,m.

Although both methods are extremely sensitive to very small gradients in the electron density, they are both most sensitive to structures larger than $\sim$1\,km.
Since refractive shifts are measured in the image plane, they are averaged over all baselines, which for most visibility weighing schemes leads to an average baseline length $>$1\,km for the MWA.
For measurements based on phase calibration solutions, the relatively small phase variations due the ionosphere on shorter baseline lengths are small compared with in phase variance due to thermal noise for baselines shorter than a few hundred metres \citep[see e.g. fig. 3 of][]{mevius2016}.
The smallest ionospheric structure may also be undetectable to both methods if the data are integrated too much in time, since even with very low drift velocities of $\sim$10\,m/s  \citep[e.g.][]{asaki2007} smaller structures will drift over the array in less than a minute.

\begin{figure}
    \centering
    \includegraphics[width=0.92\textwidth]{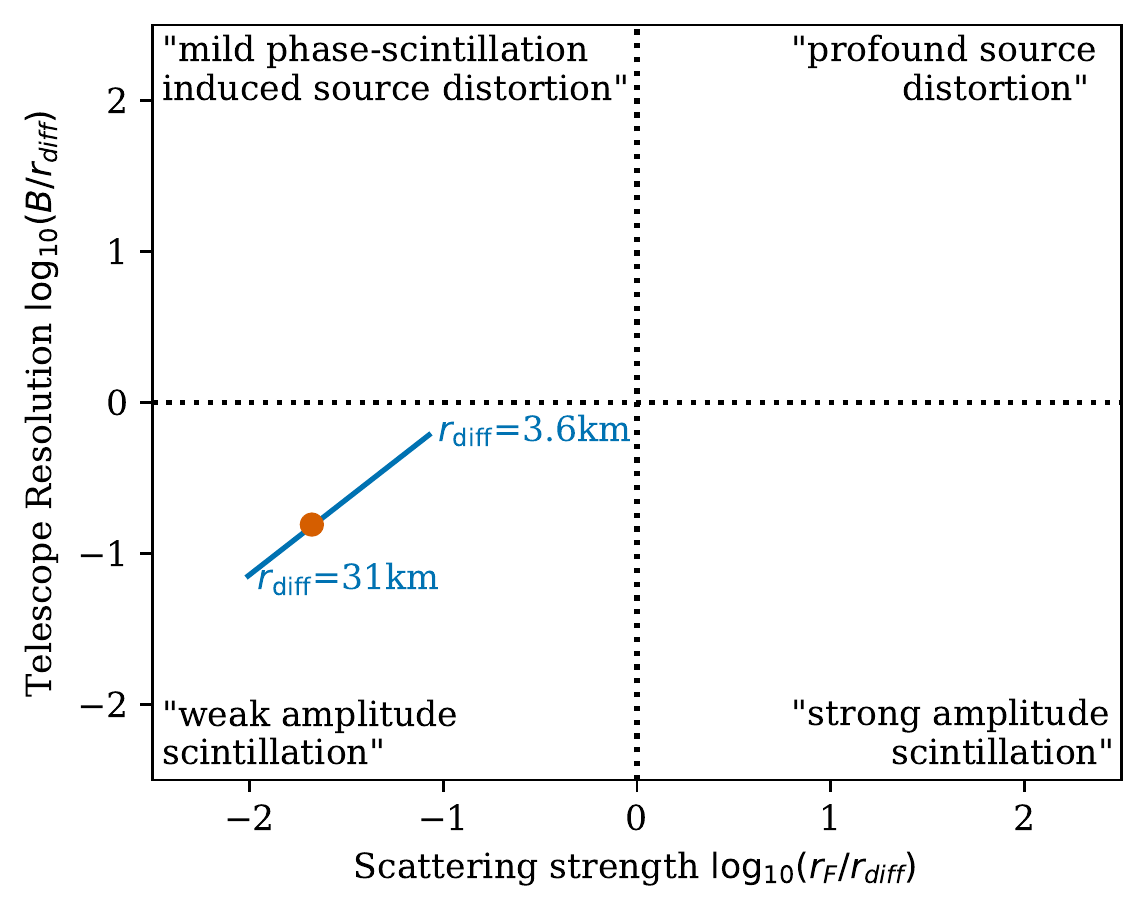}
    \caption[MWA Scintillation Regimes]{Scattering strength vs Baseline length in terms of scattering scale, following \cite{cornwell1989}. Descriptions of asymptotic regimes are also from \citeauthor{cornwell1989} (though we adopt our definition of $r_F$; see \eqn~\ref{eqn:r_f}). The blue line is for the range of $r_\mathrm{diff}$ observed by \citet{mevius2016}, but scaled to our observing frequency of 154\,MHz. Baseline length $B$ is assumed to be 2.2\,km (MWA Phase I). the orange point calculated from the scintillation index observed by \citep{morgan2021} with the MWA at 162\,MHz. Height of the ionosphere is assumed to be 300\,km.}
    \label{fig:cornwell}
\end{figure}

The fact that the ionosphere at the small, $\sim$300\,m scales remains unexplored with the MWA motivated us to investigate the use of ionospheric scintillation to probe these scales.
The effects of ionospheric scintillation on Global Navigation Satellite Systems (GNSS) is well-known, and is a widely-used probe of the ionosphere \citep{kintner2007} in the more extreme ionospheric conditions found in the polar \citep{spogli2009} and equatorial regions \citep{beach1997}. 
Scintillation increases with observing wavelength.
Therefore, at MWA and LOFAR frequencies, an order of magnitude lower than those used for GNSS, we would expect scintillation to be measurable even for the benign mid-latitude ionosphere observable where these telescopes are (by design) located.

Depending on the observing parameters, scattering effects can manifest in different ways, including amplitude scintillation as well as rapid image-plane distortion.
This is illustrated in \Fig~\ref{fig:cornwell}.
This figure shows that if we assume that the diffractive scales ($r_\textrm{diff}$) measured by  \citet{jordan2017} and \citet{mevius2016} apply on the smaller scales probed by scintillation,  we would expect to be firmly in the ``weak amplitude scintillation'' regime.
In this weak scatter regime, the scattering strength corresponds to the scintillation index, so we would expect indices of a few percent\footnote{The reader is referred to \citep{cornwell1989} for an explanation of the other regimes, which are not relevant to this paper.}. 
This is confirmed in real data from the MWA, where temporal power spectra of bright sources observed with the MWA clearly show amplitude scintillation of a few percent \citep{morgan2021}.
Dynamic spectra of a scintillating source observed by LOFAR at observing frequencies of 10--80\,MHz by \citet{2020JSWSC..10...10F} are also consistent with this picture, with weak amplitude scintillation in the upper half of the band, and strong amplitude scintillation (characterised by spectral structure) evident at the lowest frequencies.

Weak scintillation measurements are sensitive to structures close to the Fresnel Scale $r_\text{F}$ (defined in \eqn~\ref{eqn:r_f}), around 300\,m at MWA frequencies and typical ionospheric heights, thus allowing us to probe spatial scales an order of magnitude smaller than those probed by refractive shifts, and even smaller than the scales probed by \citet{2022JATIS...8a1012R}.

Beyond the opportunity to explore new parameter space, amplitude scintillation is also interesting in that it may be measured with much more modest instruments than LOFAR or the MWA, since it does not require baselines longer than 100\,m.

Below, we present amplitude scintillation measurements from several thousand MWA Epoch of Reionisation observations \citep{trott2020}. Importantly, the ionosphere has already been characterised via refractive shifts for all of these observations, permitting a comparison of these two observables.

The paper is organised as follows: in section~\ref{sect:observations} we describe our observations and the refractive shift analysis that has already been completed.
In section~\ref{sect:method} we describe our methodology for measuring ionospheric scintillation.
In section~\ref{sect:results} we present our results and in section~\ref{sect:discussion} we give our conclusions and offer our suggestions for future work.

\section{OBSERVATIONS}
\label{sect:observations}
As an extension of the work in \citet{jordan2017}, more EoR observations were calibrated and analysed for their ionospheric activity.
While the results of this work have not been published in their own right, they have still contributed to other projects, such as that presented in \citet{trott2020}.

In total the ionospheric refractive shift analysis has been carried out on 29\,070$\times$2-minute observations across 350 nights.
These nights include data from the `EoR-0', `EoR-1' and `EoR-2' MWA fields.
All observations were taken between August 2013 and January 2016 in the MWA Phase I configuration, and were conducted in the ``drift-and-shift'' strategy where the pointing centre is periodically updated to keep the target EoR field close to the centre of the primary beam.
They utilise the full 30.72\,MHz of MWA instantaneous bandwidth in one contiguous block at one of two centre frequencies: 182\,MHz (EoR ``high-band'': 10\,777 observations) or 154\,MHz (EoR ``low-band'': 15\,425 observations).

For this work we restricted ourselves to Low-band EoR-0 observations: which covers 5\,251 observations in total.
However, visibility data for some of these observations are no longer available (typically ones observed further from the zenith).
We also restricted ourselves to observations taken in 2014, and rejected those with 8 or more tiles flagged.
This left us with 2\,070 observations for further analysis.

\begin{figure}
    \centering
    \includegraphics[width=0.95\textwidth]{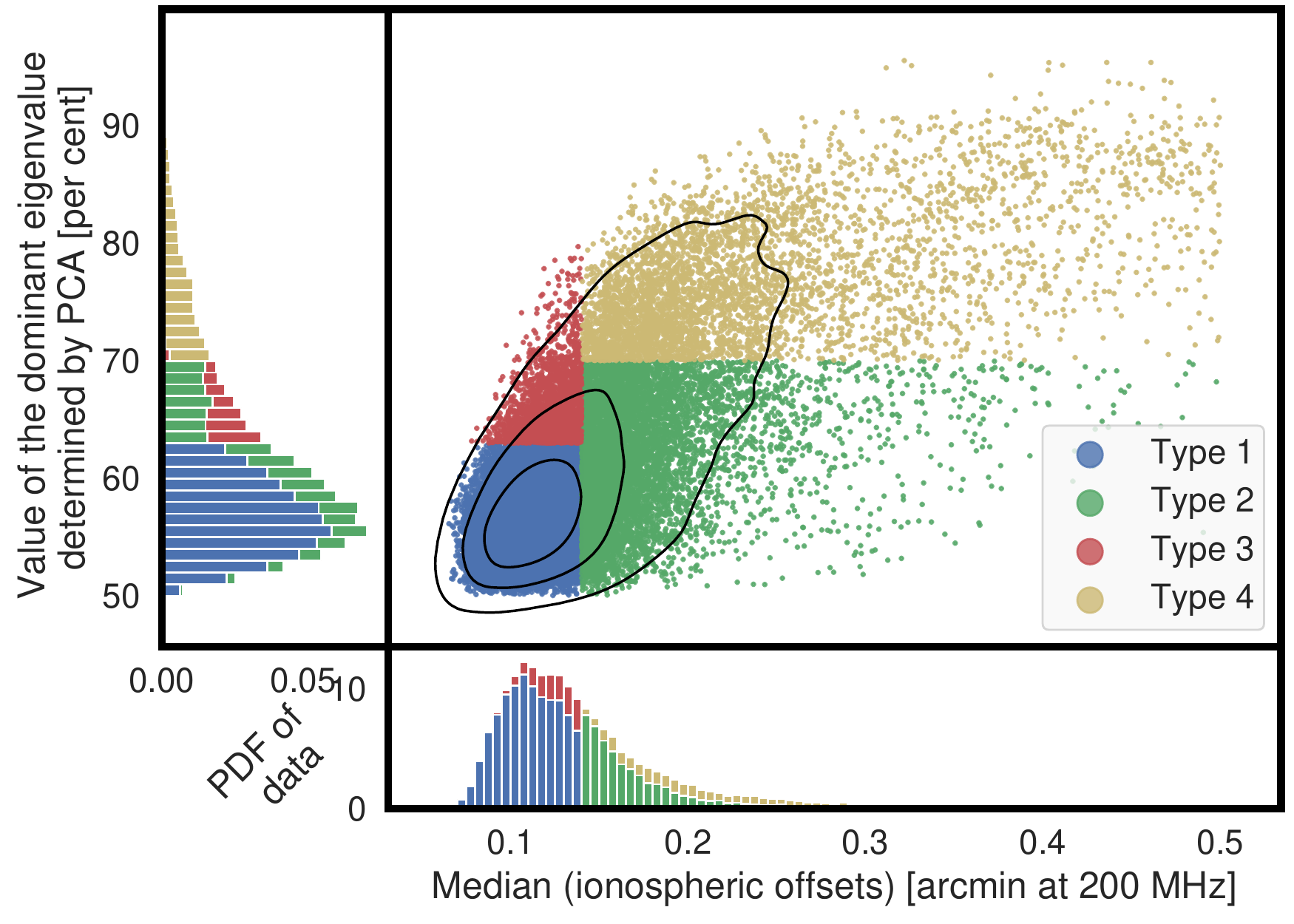}
    \caption[Scatter of Ionospheric Type Metrics]{A scatter plot of the median ionospheric offset versus the anisotropy, which is the dominant eigenvalue determined by PCA for all 29\,070 EoR MWA observations. The 4 distinct ionospheric populations, or 'types', are also included.}
    \label{fig:metric_scatter}
\end{figure}

\subsection{REFRACTIVE SHIFT ANALYSIS}
As with \citet{jordan2017}, these 29\,070 observations have had their ionospheric activity categorised into `magnitude' and `anisotropy' components.
The `magnitude' corresponds to the median observed ionospheric refractive shift, after they have been normalised to 200\,MHz (ionospheric shifts scale with the square of wavelength), and the `anisotropy' is determined by a Principal Component Analysis (PCA).
The ionospheric shifts are two-dimensional (typically in Right Ascension and Declination) and so a PCA yields two eigenvalues; if the eigenvalues are the same, the data is purely anisotropic.
On the other hand, if there is even a slight trend for ionospheric shifts lie along an axis, one eigenvalue will be larger the other; it is the dominant eigenvalue that is recorded for categorisation, normalised such that the sum of the eigenvalues is unity (i.e. the anisotropy values all lie in the range 0.5--1.0).

Using the ionospheric data collected from each observation, \citet{jordan2017} categorised four populations of ionospheric activity, as shown on \fig\,\ref{fig:metric_scatter}. These classifications were labelled as Types 1 through 4, all of which have been specifically identified using the following criteria:
\begin{equation*}
    m < 0.14 \text{ and } p < 63 \Rightarrow \text{Type 1,}
\end{equation*}
\begin{equation*}
    m > 0.14 \text{ and } p < 70 \Rightarrow \text{Type 2,}
\end{equation*}
\begin{equation*}
    m < 0.14 \text{ and } p > 63 \Rightarrow \text{Type 3,}
\end{equation*}
\begin{equation*}
    m > 0.14 \text{ and } p > 70 \Rightarrow \text{Type 4,}
\end{equation*}
where $m$ is the median ionospheric offset and $p$ is the dominant eigenvalue in per cent.
This categorisation has divided the total 2\,070 observations in this analysis into approximately 70, 6, 13 and 11 per cent for ionospheric Types 1 through 4, respectively. 

\subsection{IPS SOURCES IN THE EOR-0 FIELD}
Interplanetary Scintillation (IPS) causes compact (sub-arcsecond) sources to vary in brightness on $\sim$1\,s timescales.
IPS on the nightside is weaker than it is closer to the Sun; nonetheless scintillation indices can be as high as 5\% under typical conditions \citep{Bell:phdthesis}, and may occasionally be much higher \citep{2015ApJ...809L..12K}.
Although somewhat separable from ionospheric scintillation due to its shorter timescale, IPS is nonetheless an important potential contaminant.
However, in contrast to ionospheric scintillation, which will affect all discrete sources in the EoR field, only the very compact components of sources will show IPS.
We therefore utilised as-yet unpublished MWA IPS observations covering the EoR-0 field from the MWA IPS Survey \citep{2019PASA...36....2M} to determine the extent to which each source in the field might be affected by IPS.
Restricting ourselves to the brightest sources \citep{White}, for each we calculated the Normalised Scintillation Index \citep[NSI; ][]{2018MNRAS.474.4937C}: the ratio of the observed scintillation index to that expected for a point source.

These NSIs are listed in \tab~\ref{tab:ips}.
Errors are approximately 20\% (note also that since the IPS observations were made when the Sun was active, it is possible that some of these NSIs are overestimated).
By selecting only NSI$<0.2$ sources and filtering out IPS timescales (see \sect~\ref{sec:scint}), we reduce IPS variability to well below 1\%.

\section{METHOD}
\label{sect:method}

\subsection{Calibration}
First, a calibration solution was produced for each observation \citep{2015PASA...32....8O} using a sky model \citep{gleam2017} attenuated by the primary beam.
A single calibration solution was produced using the full two-minute observation.
In addition to this, purely for quality control purposes, we found it useful to generate a calibration solution for each individual 2-s correlator integration time.
Examination of the phase as a function of time using these calibration solutions allowed us to identify phase jumps due to instrumental issues, which affected a small subset of observations.

\subsection{Imaging}
\begin{figure*}
    \centering
    \includegraphics[width=0.9\textwidth]{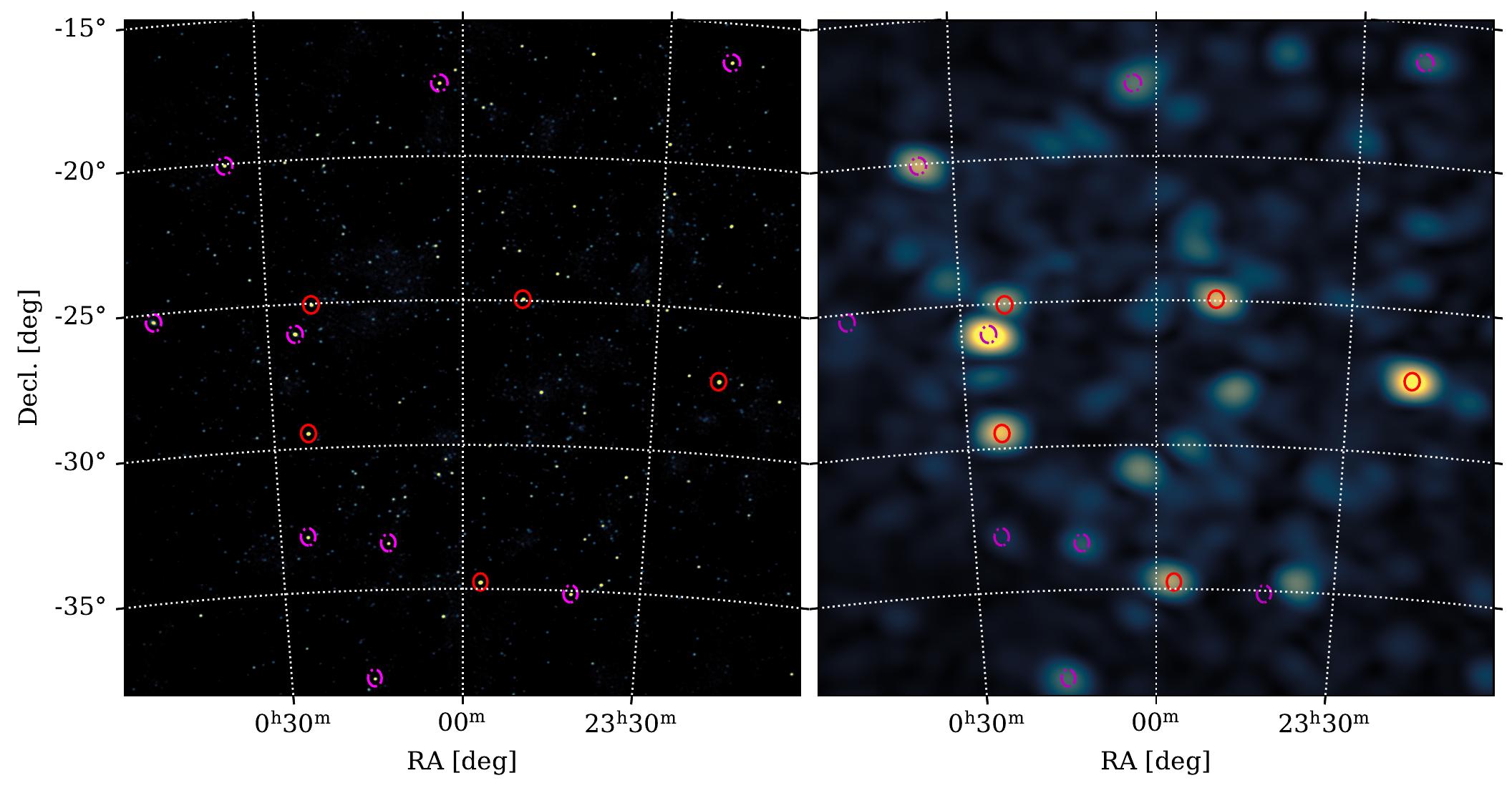}
    \caption[Standard vs Variability Image]{Standard image (left) and variability image (right) of the centre of the EoR-0 field for a strong scintillation observation (18.1\%). Indicated are the top 15 brightest sources that are catalogued in the GLEAM 4Jy sample \citep{White}, with the high-NSI sources (magenta) distinguished from the 5 low-NSI sources (red) which are the ones used to calculate the scintillation index for an observation.}
    \label{fig:eor0field}
\end{figure*}

By applying the calibration solutions and using WSClean \citep{offringa2014wsclean}, a `standard image' was created for each observation using the whole 2 minutes of observing time, and all unflagged tiles of the full Phase I MWA.
This image was deconvolved using image-based CLEAN interspersed with up to 5 ``major cycles'', where the CLEAN components were subtracted from the visibilities.
The final `model' visibilities, comprising all the CLEAN components, was written to the visibility measurement set for later use.
A final standard image, as shown on the left in \fig \ref{fig:eor0field}, was created by combining the individual synthesis images of both linear polarisations XX and YY with appropriate weightings \citep[\eqn~1]{1996A&AS..120..375S}.

Next, `snapshot' images were generated for each individual 2-s time integration for each polarisation (excluding the first 2 and last 3 integrations, resulting in 51 snapshots per observation).
These differed from the standard images significantly in two ways: first, the model derived from the standard image deconvolution process was subtracted from the visibilities, leaving only the residuals to be imaged.
Second, any tiles from beyond the central 100-m `core' of the MWA were flagged, reducing the number of tiles to 36 and decreasing the average baseline length to smaller than the Fresnel scale ($r_\text{F}\sim$300\,m), whereas the resolution of the snapshot images (1.38\degr$\times$1.06\degr) corresponds to an average baseline length $\sim$100\,m.

The snapshot images were then stored using the image cube format described by \citet{morgan2018} Appendix I.
This image format allows for the examination of the time series corresponding to any pixel, as well as facilitating the other timeseries analyses described below.

\begin{figure*}[h!]
    \centering
    \includegraphics[width=0.47\textwidth]{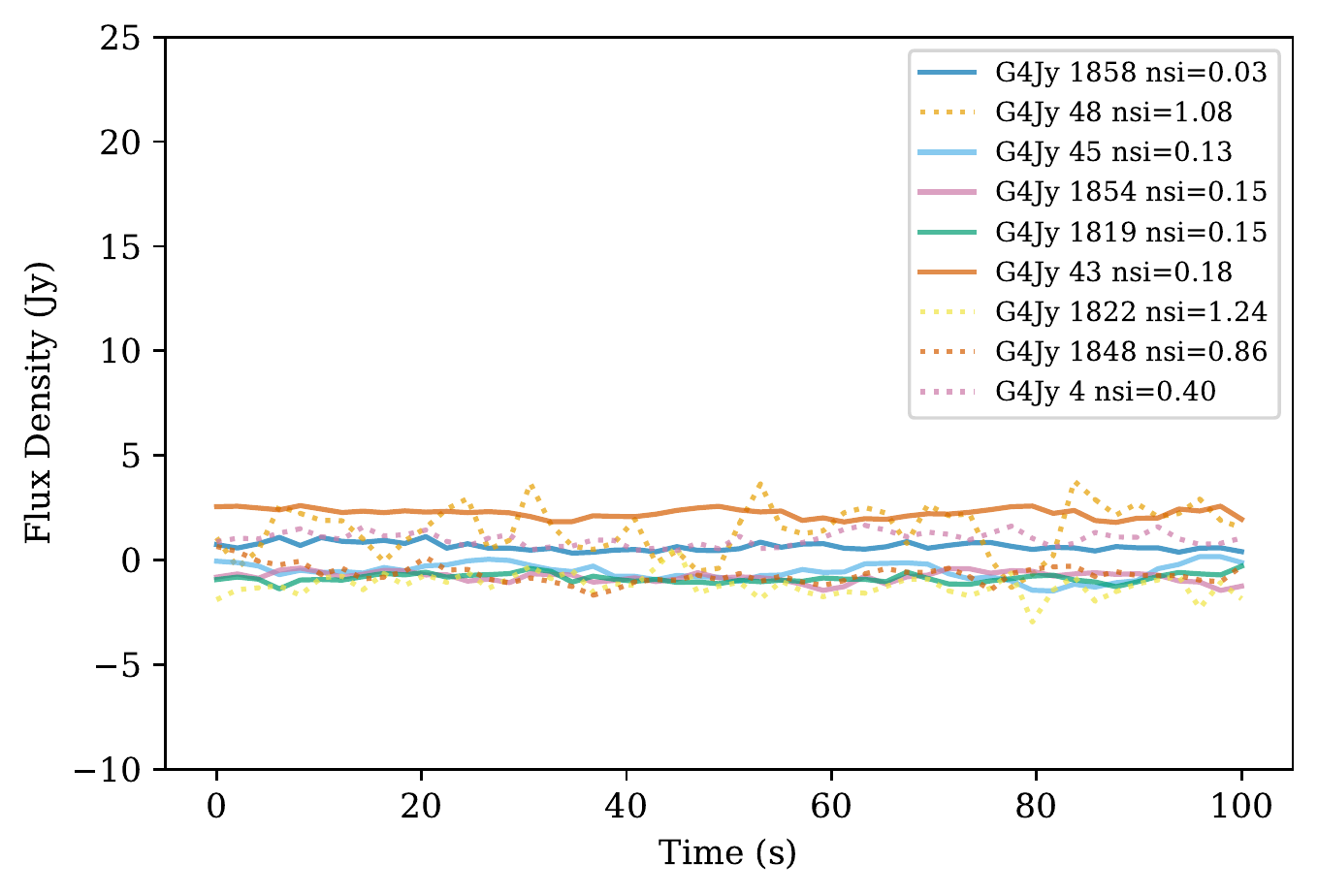}\hspace{0.7cm} \includegraphics[width=0.47\textwidth]{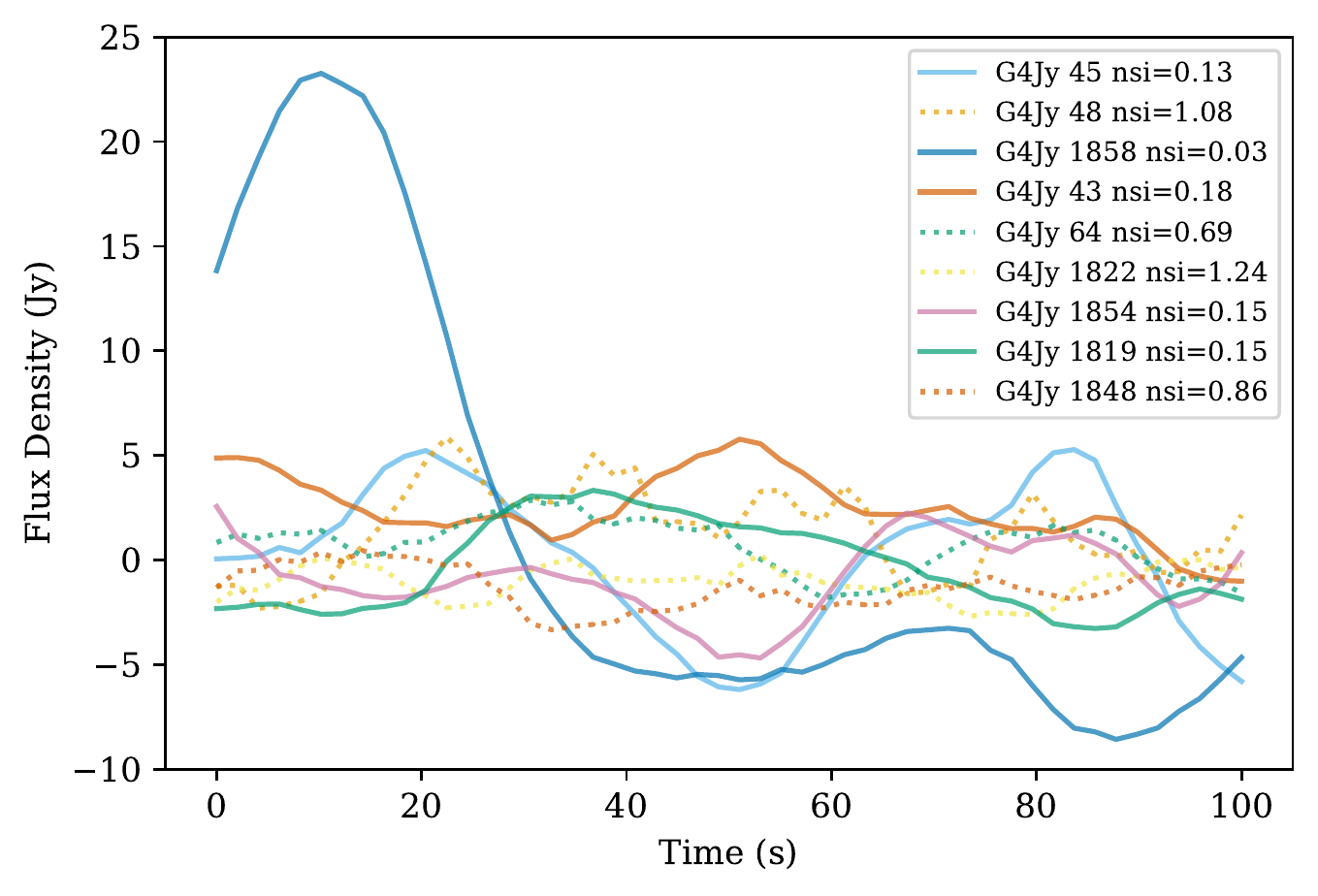}
    \caption[Timeseries]{Time series analysis for low- (left, 1.2\%) and a high- (right, 24.8\%) scintillation observation, both at a frequency of 154MHz, with off-source noise included. Sources with an NSI of above 0.2 are classified as high-NSI, or strong-IPS sources and are shown as dotted lines. These sources will follow variability on IPS timescales (1-2s), while the low-NSI, weak-IPS sources (filled line) display variation on a timescale of 10s, which is consistent with ionospheric scintillation, particularly prominent in higher scintillation observations.}
    \label{fig:timeseries}
\end{figure*}
\Fig~\ref{fig:timeseries} shows timeseries for the brightest sources in the field of view for observations representative of both `high' and `low' ionospheric scintillation.
Variability on timescales consistent with ionospheric scintillation (which is clearly resolved with time resolution of 2\,s) is clearly visible, as is IPS for those sources with NSI$>$0.2.
Note that the timeseries have a non-zero mean offset. This may be due to the presence of extended structure in the field \citep{lenc2017} which will not be included in the subtracted model.
The observation length is relatively short compared to the scintillation timescale, so averaging over sources and/or observations will be necessary in order to avoid sampling errors.

\begin{figure}
    \centering
    \includegraphics[width=\textwidth]{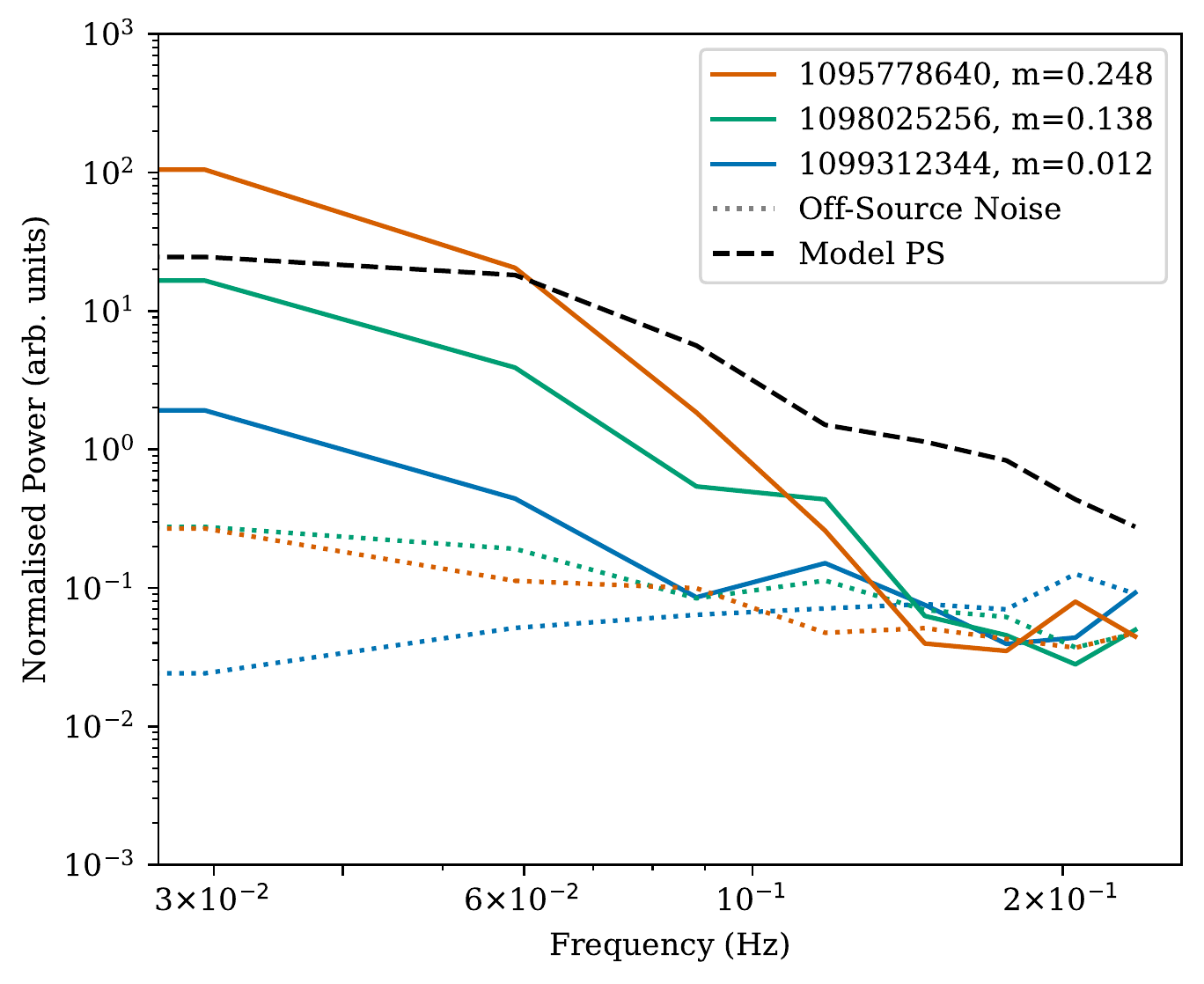}
    \caption[Power spectra - G4Jy 45]{Power spectrum analysis for a bright, low-NSI source, G4Jy 45 (GLEAM J002430-292847) for a high (24.8\%), mid (13.8\%), and low (1.2\%) scintillation observation, with the off-source noise power spectra included. 
    The presence of the `Fresnel knee' at the corresponding frequency assures ionospheric scintillation is being analysed, and each observation drops off to the background noise level at higher frequencies.
    Included in the figure is a model power spectrum weak scintillation theory \citep{macquart2007}, with a height to the ionosphere taken as 300km \citep{hunsucker_hargreaves_2002, wild1956}, and a velocity of 50m/s \citep{asaki2007}.}
    \label{fig:powerspectra}
\end{figure}
\Fig~\ref{fig:powerspectra} shows power spectra for a specific source in high, medium and weak scintillation observations.
The power spectrum shows a shape characteristic of scintillation, with high low-frequency power and the Fresnel `Knee' marking the onset of a power-law drop-off down to the noise level.

\subsection{Scintillation Analysis}
\label{sec:scint}
Following \citet{morgan2018}, it is possible to create an image of the variability in the field of view by taking the standard deviation of the flux density time series after filtering the timeseries to emphasise timescales of interest.
This consisted of applying a low-pass Butterworth filter (order 2) with a cutoff frequency of 0.125\,Hz to attenuate IPS variability.
This `variability' image, as seen on the right of \fig \ref{fig:eor0field}, is a summary of the time series as it collapses the time dimension of the previously created three-dimensional image cube.
All pixels will have positive values (which are non-zero due to thermal noise).
Any pixel corresponding to a source undergoing amplitude scintillation will have excess variance, and it is possible to identify sources with significant excess variability as
\begin{equation}
    \label{eqn:p_mu_sigma}
    P - \mu > 5\sigma
\end{equation}
where $P$ is the pixel value, $\mu$ is the background level, and $\sigma$ is the spatial rms of the image.
Identification of varying sources is therefore very similar to the problem  of identifying discrete sources in any astrophysical image.

After the creation of the variability images, Aegean \citep{hancock2012} was used to identify sources in the standard image.
Once the sources were identified, they were cross-matched with the GLEAM 4Jy sample \citep{White}: a subset of the MWA GLEAM survey \citep{gleam2017}.
The brightest non-IPS sources (NSI$<$0.2) were then identified (usually the 5 `Low-NSI' sources shown in \fig~\ref{fig:eor0field}).
The pixel in the moment image closest to the continuum detection was measured as well as four off-source pixels (25 pixels away) surrounding the detection.

It is then possible to calculate the scintillation index $m$ of each source
\begin{equation}
    m = \frac{\Delta S}{S_\text{standard}}
\end{equation}
where $S_\mathrm{standard}$ is the brightness of the given source as is found in the standard image and $\Delta S$ is the excess variability of the source.
Since the variability image values are standard deviations (i.e. the square root of the variance), the noise must be subtracted in quadrature. Therefore, 
\begin{equation}
    \label{eqn:ds}
    \Delta S^2 = P^2 - \mu^2
\end{equation}
where $P$ and $\mu$ are as defined in \eqn~\ref{eqn:p_mu_sigma}.

The median scintillation index across all sources was then computed for each observation, reducing the scintillation level of each observation down to a single number which can be compared with metrics derived from refractive shifts.

\section{RESULTS}
\label{sect:results}
\begin{figure*}[h!]
    \centering
    \includegraphics[width=0.49\textwidth]{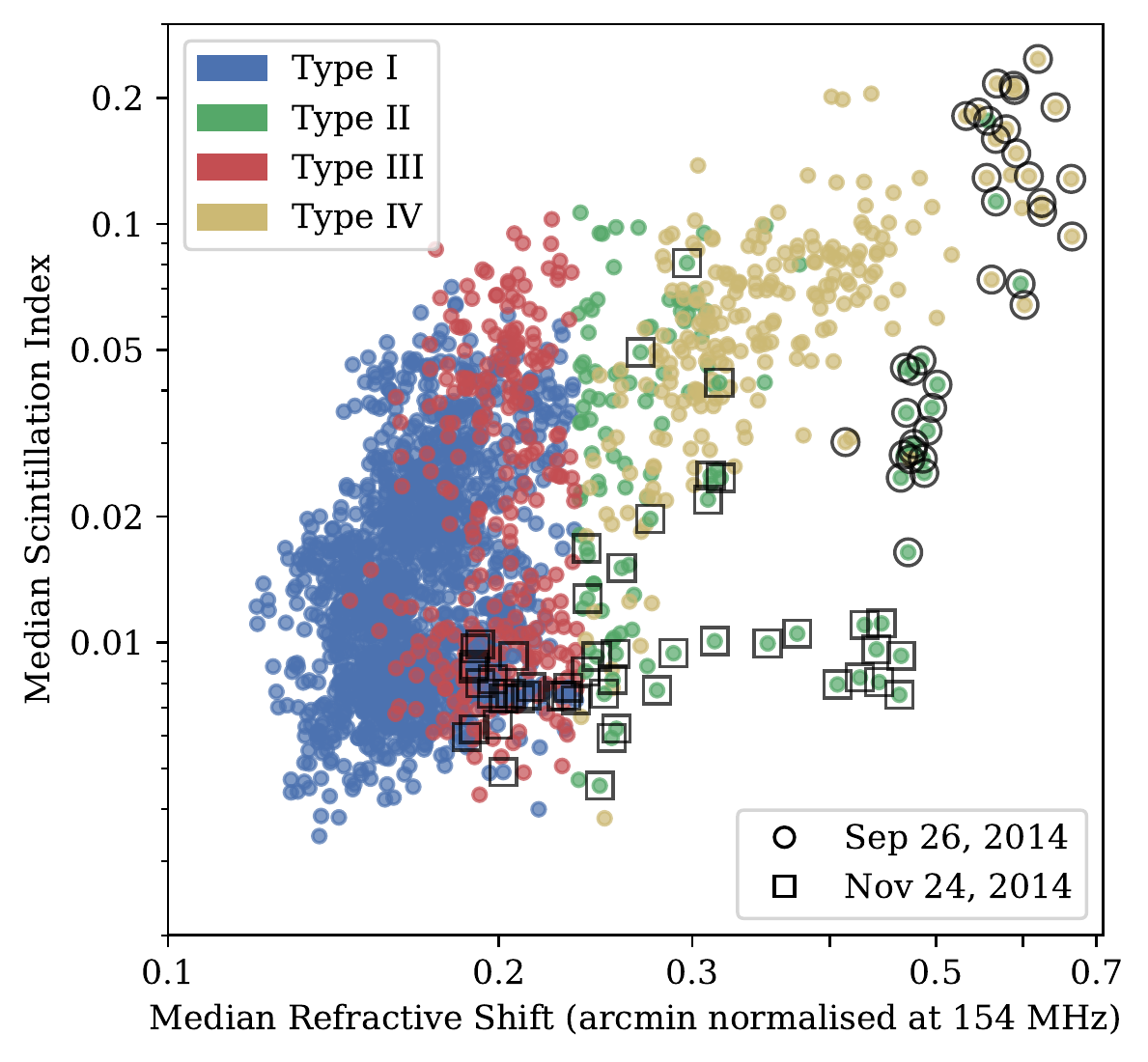}
    \includegraphics[width=0.498\textwidth]{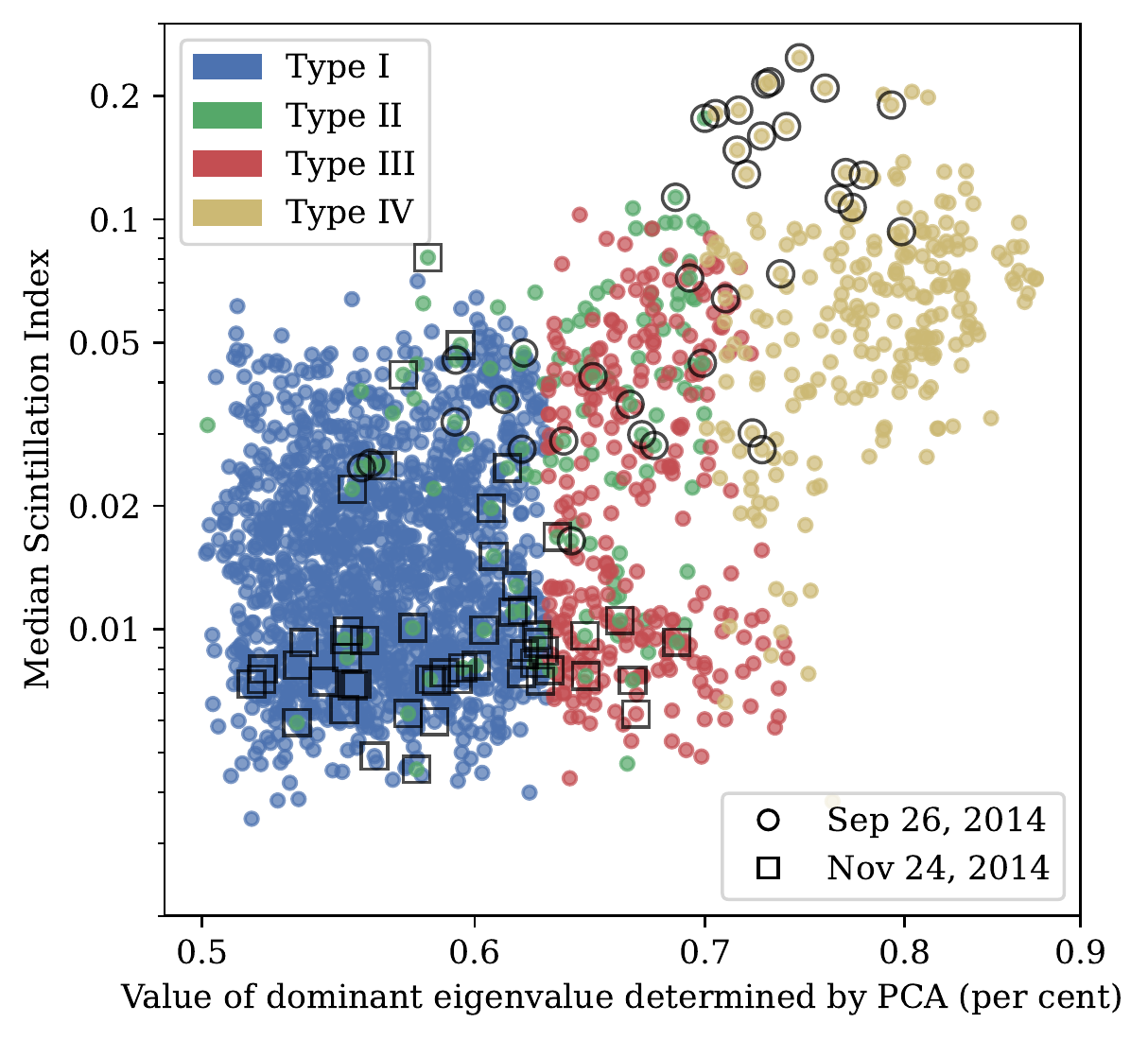}
    \caption[Scintillation index compared existing ionospheric metrics]{The median scintillation index calculated using the low-NSI sources in the MWA EoR-0 field, compared to the median refractive shift (left) and the anisotropy of an observation, a value of the dominant eigenvalue as determined by a principal component analysis (PCA) (right). The refractive shift as found by \citet{jordan2017} is calculated at a frequency of 200MHz, therefore for this comparison the refractive shifts were normalised to a frequency of 154MHz.
    Highlighted in both figures are observations from two nights of particular interest; 2014-11-24 and 2014-09-26, both of which have a Travelling Ionospheric Disturbance (TID) traverse the field.}
    \label{fig:magcompare}
\end{figure*}

The left panel of \fig~\ref{fig:magcompare} compares the calculated scintillation index and the positional refractive shifts for all observations analysed.
There is clearly a strong, positive correlation between the two metrics (Pearson correlation coefficient 0.71).
However, the correlation breaks down for some observations.
In particular there is a small subset of observations with a much higher refractive shift / lower scintillation index than the general trend.
All of these outliers belong to two nights, as indicated in \fig\,\ref{fig:magcompare}.
These two nights are discussed further in \sect~\ref{sec:outliers} below. 

There is also a less extreme set of observations where the observed scintillation index is higher than the trend.
The refractive shift of these sources spans all but the very lowest refractive shift.

The existence of any correlation between the anisotropy metric, either positive or negative, is far less clear: the apparent correlation between scintillation index and anisotropy in the right panel of \fig~\ref{fig:magcompare} (see also \fig~\ref{fig:progression} which shows the anisotropy as a colour bar) appears to be driven more by the correlation between refractive shift and anisotropy.

\subsection{DISCREPANT NIGHTS}
\label{sec:outliers}
As already noted, almost all the most discrepant points in the plot of refractive shift vs scintillation index belong to two specific nights: 2014-09-26 and 2014-11-24. On both these nights, a handful of observations were classified as Type II's, although they are all very close to the Type II / Type IV boundary, indicating the presence of higher levels of structure.  

\begin{figure*}
    \centering
    \includegraphics[width=0.98\textwidth]{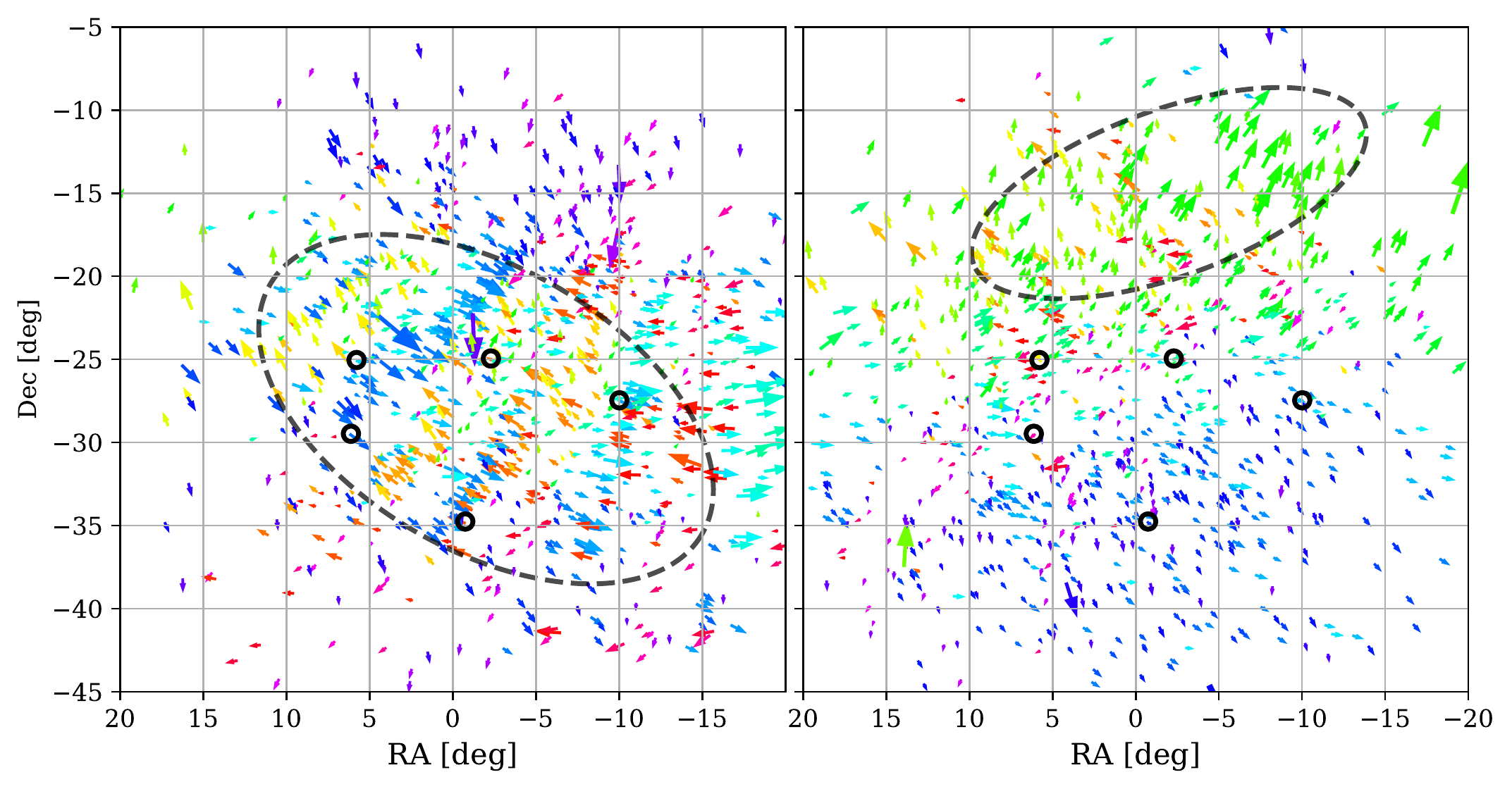}
    \vfill
    \includegraphics[width=0.98\textwidth]{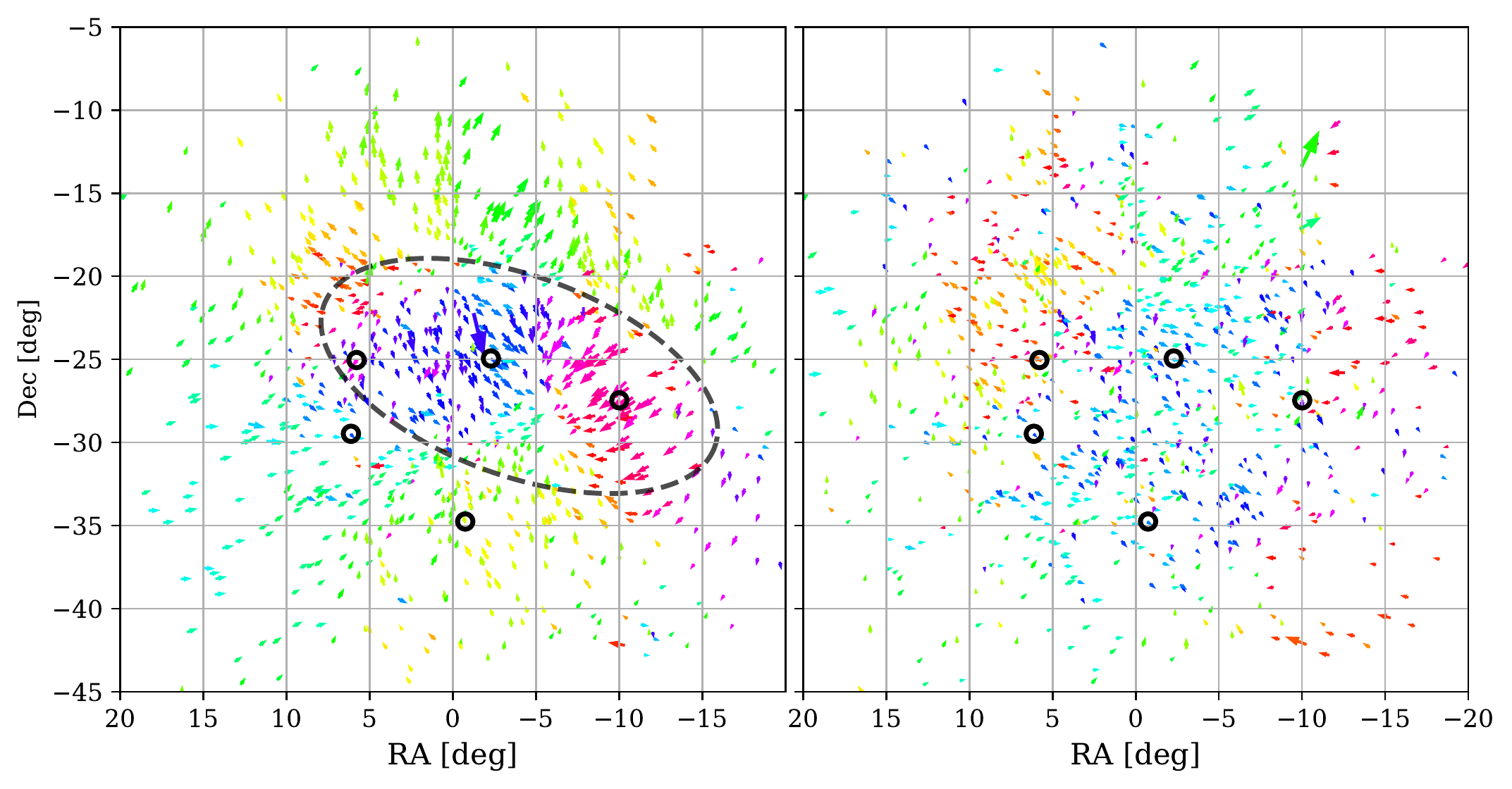}
    \caption[TID traversing EoR-0 Fields]{Plots of refractive shifts as measured by \citet{jordan2017} with each arrow representing the refractive shift of a source (arrows are also coloured by direction to emphasise structure). Circles indicate sources used for scintillation analysis. Two observations are shown for each of two TID events that were recorded in the data set, where the one above is the event from 2014-09-26 and below is from 2014-11-24. For each night, two observations are displayed; an observation from the middle of the night on the left, showing the TID at it's most central and maximum intensity, and an observation from the end of the night on the right, once the TID has moved along. The ellipse marks the approximate location of the observed TIDs. For both events, the TID is clearly visible in the centre of the field on the leftmost plot. For the event of 2014-09-26, the TID continues to persist in the field, although it has moved upwards, and is no longer being sampled by the scintillating sources. For the event of 2014-11-24, the TID completely moves from the field by the end of the observation session.}
    \label{fig:tid}
\end{figure*}

On both of these nights a Travelling Ionospheric Disturbance (TID) traverses the field, with an exceptionally active one on 2014-09-26, leading to much higher scintillation indices in parts of the field.
As is shown on the leftmost top plot in \fig~\ref{fig:tid}, earlier in the TID's path it is directly passing over and affecting the sources that were sampled for the scintillation analysis, but as it moves north, shown on the rightmost top plot in \fig~\ref{fig:tid}, the sources that were being monitored go back to lower levels of scintillation, whereas due to larger sampling used, the TID was still being captured by the shift metric. 
Thus, the apparent discrepancy between the two metrics on 2014-09-26 is due to sampling differences.

\begin{figure}
    \centering
    \includegraphics[width=\textwidth]{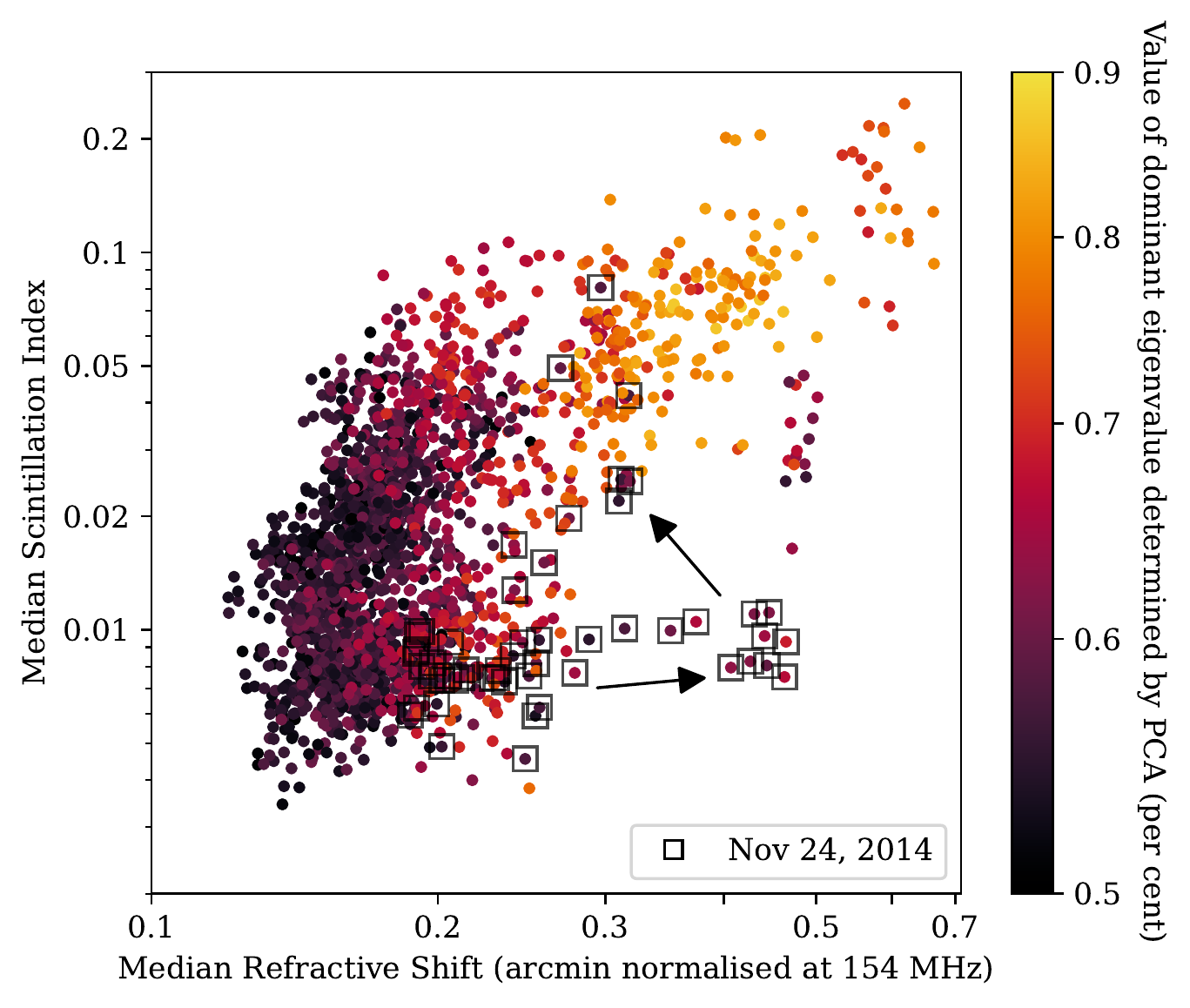}
    \caption[Scintillation Index compared to the shifts with PCA]{The median scintillation index calculated using the low-NSI sources in the brightest sources in the MWA EoR-0 field compared to the median refractive shift as recorded by \citet{jordan2017}, with the level of anisotropy represented by the value of the dominant eigenvalue as determined by a principal component analysis (PCA). Arrows depict time progression of the observations throughout the night of the 2014-11-24; see \sect~\ref{sec:outliers} for further details.}
    \label{fig:progression}
\end{figure}

In contrast, the location of the TID  that was captured by observations from 2014-11-24 (lower panels of \fig~\ref{fig:tid}), is well-sampled by the scintillating sources, even where the observations are maximally discrepant as those shown in \fig~\ref{fig:magcompare}. 
The night of this TID event can be separated into three clusters, each occupying a different space within \fig~\ref{fig:progression}.
Before the TID enters the field, both metrics indicate low ionospheric activity.
Then the refractive shifts increase, then the ionospheric scintillation increases as the refractive shifts decrease.
This is indicated in \fig~\ref{fig:progression}, where the arrows between the clusters indicate the progression of time.

\subsection{RELATING REFRACTIVE SHIFT AND SCINTILLATION INDEX WITH POWER-LAW TURBULENCE}
\label{sec:turbulence}
In turbulent scattering media, the phase structure function takes the form of a power law \citep{1979RaSc...14.1135R}
\begin{equation}
    D_{\phi}(r) = \left(\frac{r}{r_{\text{diff}}}\right)^{\alpha-2}
\end{equation}
where the `diffractive scale', $r_{\text{diff}}$, is the spatial scale over which the phase variance is 1 $\text{rad}^2$ and the $\alpha$ is the power law index: 11/3 for power-law turbulence \citep{Narayan1992}.
\citet{mevius2016} found values for $\alpha$ in the range 11/3--12/3, and $r_\mathrm{diff}$ in the range 3.5--31.1 km.

The ratio of the phase structure function for different values of $r$ is given simply by 
\begin{equation}
    \label{eqn:frac_d}
    \frac{D_{\phi}(r_1)}{D_{\phi}(r_2)} = \left(\frac{r_1}{r_2}\right)^{\alpha-2}
\end{equation}
Both the scintillation index and the refractive shift are measures of the structure function on different scales.

The scintillation index, $m$, in the weak regime measures the structure function on the Fresnel scale $r_\text{F}$ \citep{Narayan1992}
\begin{equation}
    \label{eqn:m}
    D_{\phi}(r_\mathrm{F}) = m^2 = \left(\frac{r_\mathrm{F}}{r_\text{diff}}\right)^{\alpha-2},
\end{equation}
where $r_\text{F}$ depends purely on the distance to the scattering screen $h$, and the observing wavelength $\lambda$:
\begin{equation}
    \label{eqn:r_f}
    r_\text{F} = \sqrt{\frac{\lambda h}{2\pi}} .
\end{equation}

Each measurement of a refractive shift measures the spatial gradient of the phase screen.
All baselines contribute to the measurement, so if we take the average baseline length, $B$ (where $B$=2.2\,km for our data \citealp{jordan2017}),
\begin{equation}
    \label{eqn:theta}
    D_{\phi}(B) = \left(\frac{2 \pi \theta B}{\lambda}\right)^2,
\end{equation}
where $\theta$ is the refractive shift in radians.
Note that we take the median shift of all the sources in the field of view: an average of a spatial ensemble of point measurements of $D_{\phi}(B)$ over a region of the ionosphere $\sim$100\,km across \citep[c.f.][Section~4.2.1]{jordan2017}.

\begin{figure}
    \centering
    \includegraphics[width=0.95\textwidth]{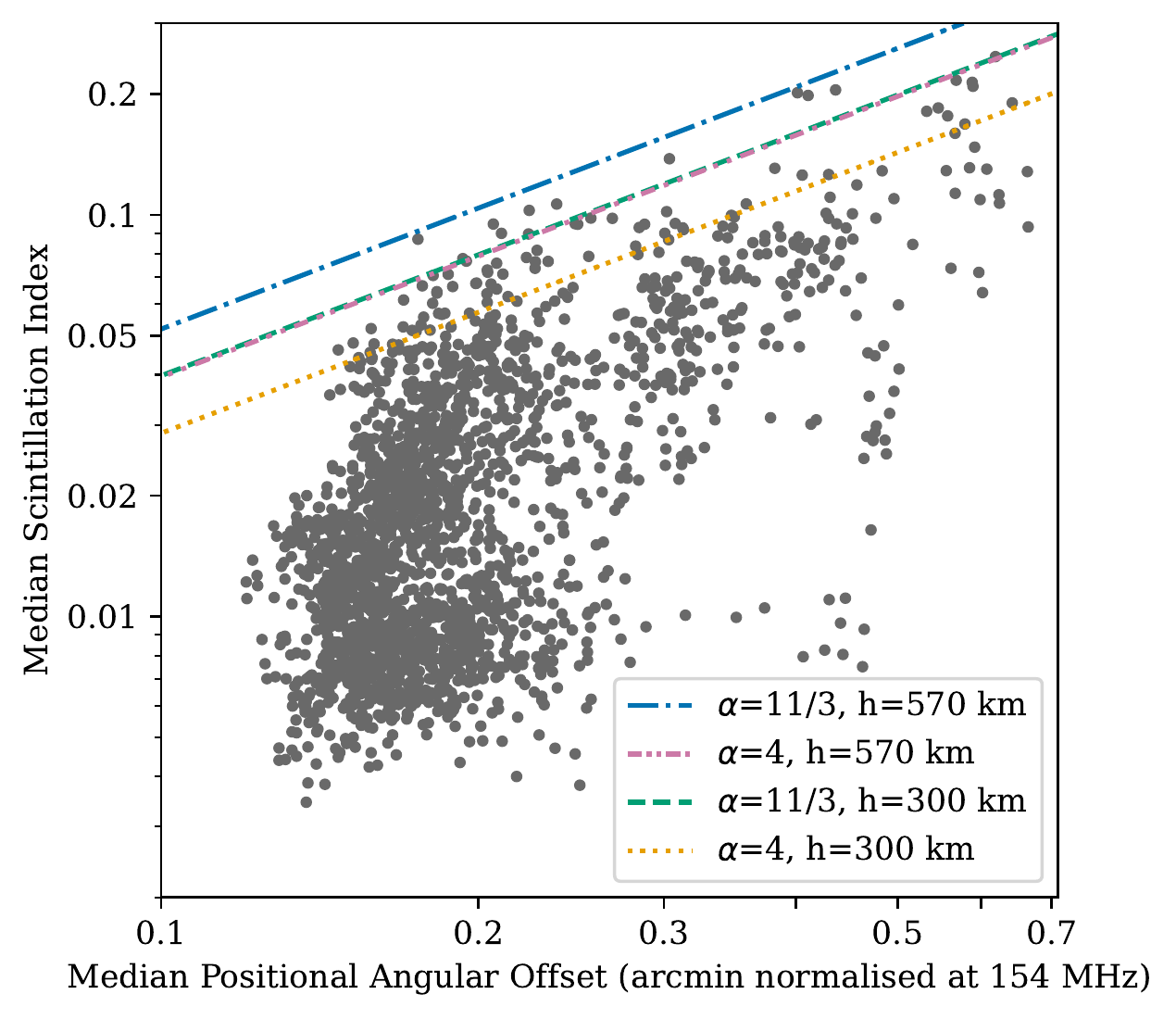}
    \caption[Scintillation Index compared to the shifts with Kolmogorov Turbulence]{Relationship between refractive shift and scintillation index assuming power-law turbulence. $11/3<\alpha\le4$ is the range of spectral indices found by \citet{mevius2016}. Lower and upper limits on the height to the ionosphere, $h$, at 300\,km \citep{hunsucker_hargreaves_2002, wild1956} and 570\,km \citep{cleo2015a}, corresponding to Fresnel scales of 305\,m and 420\,m respectively.}
    \label{fig:kolcompare}
\end{figure}
\Fig~\ref{fig:kolcompare} shows the expected relationship between $m$ and $\theta$ from \eqns~\ref{eqn:frac_d}, \ref{eqn:m} and \ref{eqn:theta}.
Since both $\theta^2$ and $m^2$ measure phase variance, $m$ and $\theta$ will be proportional to each other in the case of power law turbulence.
Adjusting either the power law index of the turbulence, $\alpha$, or the distance to the scattering screen, $h$, will change the constant of proportionality, but $h$ and $\alpha$ are degenerate as far as the relationship between $m$ and $\theta$ is concerned. 

Examining our measurements relative to the theoretical relationship, we find that Kolmogorov turbulence predicts a somewhat higher scintillation index for a given refractive shift than is observed for the vast majority of observations.
The expected relationship fits the data relatively well when the ionosphere is quite active; however for the majority of observations, where the ionosphere is quiet according to both metrics, the scintillation index is much lower than would be expected given the refractive shift.
We note that in the absence of refractive shifts, sources will still be shifted due to noise. 
Since the refractive shift metric was intended only to identify observations of high ionospheric activity, no attempt was made to subtract this noise.
It is also possible that non-turbulent structures, such as those discovered by \citet{cleo2015a} are persistent in the refractive shift data at some level.
For our ionospheric scintillation measurements, the noise \emph{was} subtracted (see \eqn~\ref{eqn:ds}), although other sources of variability, such as residual IPS, or change of instrumental response as the sky rotates may still be positively biasing the scintillation indices.
Nonetheless, it is plausible that the true refractive shifts of many of the Type I observations are much lower than those presented here, and they may therefore lie closer to the expected relationship.

\section{DISCUSSION}
\label{sect:discussion}
\subsection{Interpretation}
We have measured ionospheric scintillation across a representative sample of $2\,070\times2$-minute MWA observations and compared them with a refractive shift analysis already carried out.
The range of scintillation indices seen matches predictions reasonably well (\fig~\ref{fig:cornwell}), and there is a correlation of 0.71 between the observed scintillation index and average refractive shift.
Moreover, at least for more active ionospheres, the relationship between scintillation index and refractive shift is broadly in line with what would be expected if the ionosphere's structure is well described by Kolmogorov turbulence between the relevant scales of $2.2$\,km down to the Fresnel scale $\sim 300$\,m.
This is entirely consistent with previous observations, such as those of \citet{mevius2016}.

Power-law turbulence arises naturally when energy cascades from large scales to small scales without dissipation; the spectral index of -11/7 arising from dimensional arguments \citep{1941DoSSR..32...16K}.
Kolmogorov turbulence is an excellent model for a broad range of astrophysical plasmas \citep{rickett1990}, however departures from Kolmogorov turbulence are also common.
All turbulent media must have an outer scale (set by the dimensions of the medium if nothing else) and an inner scale.
For most ionised astrophysical plasmas (including the ionosphere) the latter is identified with the proton cyclotron scale ($\sim 5$\,m for the ionosphere \citealp{1978JATP...40..803B})

The evolution of the ionosphere in the MWA field of view over the course of the night of 2014-11-24 (\fig~\ref{fig:progression}) is consistent with energy injection on large scales (the TID), followed by a cascade down to smaller scales (scintillation index increases while the refractive shifts slightly decrease).
This would mean that our strongest departure from Kolmogorov turbulence is due to the turbulence not being fully developed at the time of observation.
However since we cannot distinguish in our data between the group velocity of the TID and any bulk flow, this interpretation is not unambiguous.

The most conspicuous outliers from the Kolmogorov turbulence model have scintillation indices well below the level predicted from the measurements at larger scales: in other words, there is more structure on the larger scales.
This is unsurprising given the regularity with which large, coherent structures which do not resemble turbulence are observed with the MWA \citep[e.g.][]{cleo2015b}.
Indeed, \citet{cleo2016b} actually observed a case where a large-scale TID formed structures on scales of 10-100\,km which persisted for at least two days; a picture fundamentally at odds with a turbulent cascade.
However, this does raise the question of why, if these regular, large-scale structures do \emph{not} imply structure on smaller scales, there is not an obvious anti-correlation between scintillation index and the anisotropy metric of \citet{jordan2017}, which aims to quantify the extent to which structure is found on a particular scale.
In particular (see \fig~\ref{fig:progression}), observations with refractive shifts in the range 0.3\arcmin--0.5\arcmin, where the scintillation indices are almost bimodal, it is the observations which obey the Kolmogorov relation that have the highest anisotropy values.
Observations with extremely high scintillation indices ($m>0.1$) are very highly structured according to the anisotropy metric (although the distribution of anisotropy as a function of refractive shift for observations with $m>0.1$ appears identical to that for observations with $0.05<m\le0.1$).

This apparent inconsistency may simply be an artefact of what the anisotropy metric is actually measuring.
The left panel of \fig~\ref{fig:tid} shows the spatial distribution of refractive shifts during an observation where the refractive shifts are large, whereas the scintillation index (and anisotropy metric) are relatively small.
The structure in the refractive shifts is characterised less by a preferred direction (which is what the anisotropy metric measures) and more by a lack of small-scale structures (shifts of nearby sources are strongly correlated).
We therefore speculate that a different measurement of structure which measures the scale over which neighbouring sources are strongly correlated may show a clearer anticorrelation with scintillation index for a given level of ionospheric activity.

For the most extreme ionospheres observed here (refractive shift > 0.4\arcmin), the Kolmogorov spectrum occasionally underestimates the scintillation index by a factor of two or more (for a scattering height of 300\,km).
It may be that in these most extreme cases, the refractive shift for each source is actually varying over the 2-minute observation, leading to an underestimate.

\subsection{Implications for low-frequency Radio Astronomy}
In measuring the very small scales of the ionosphere, we are probing the features which have the potential to have the most pernicious effects on radio astronomy. 
The presence of relatively large-scale structures revealed by refractive shifts 
places MWA calibrations in regime 2 as defined by \citet{lonsdale2004} \citep[see also][]{2010ISPM...27...30W}, where the ionospheric irregularities are small compared to the field of view of the instrument, but large compared to the extent of the instrument.
This means that corrections can be made in the image plane \citep{2018A&C....25...94H}; although calibration and imaging algorithms also exist which allow direction-dependent corrections to be made while gridding the visibilities relatively efficiently, even when they vary quickly with time \citep{2018A&A...616A..27V}.

On the other hand, significant structure on $\sim$300\,m scales (corresponding to an angular scale of a few arcminutes at typical ionosphere heights) would place calibration very firmly in a regime where the ionospheric phase would vary not just across the field of view, but across different parts of the array \citep[][regime 4]{lonsdale2004}.
The ``curse of dimensionality'' across pointing direction, array element location, and (with a modest drift velocity) time, means that the number of independent ionospheric complex gains (amplitude and phase) rapidly accumulates and may even outstrip the number of measurements; although the number of degrees of freedom may be reduced for a sufficiently dense observing array if assumptions can be made that the ionosphere is a thin and/or frozen screen, or has non-stochastic structure \citep{2005ASPC..345..317E,2010ISPM...27...30W}.

Fortunately, the most extreme scintillation index in our work, 25\%, implies a phase variance on the Fresnel scale of less than 0.1\,rad$^2$ (\eqn~\ref{eqn:m}) which is significant, but would have limited effects in the image plane except for strong sources. Nonetheless, more extended climateology of the small-scale structure in the ionosphere above the Murchison Radio Observatory, as well as more detailed observations to characterise its properties, may be useful in assessing the impact of these irregularities on future instruments such as the SKA-low and guiding future mitigation strategies.

\subsection{FUTURE WORK}
We have limited our analysis to a straightforward comparison of median ionospheric scintillation index to existing per-observations metrics of ionospheric activity derived from the same data.
Now that variability images have been generated for all these observations, more in-depth investigations would be possible using these data, such as breaking the field of view into subfields for a finer comparison of the refractive shifts and scintillation indices. 
The refractive shift data, in particular, is an extremely rich dataset which we have averaged down to two numbers per observation.
If the distance to the irregularities is known (which can be determined by parallax: \citealp{cleo2015a}) then these 2D fields of ionospheric gradients allow the structure function of the ionosphere to be determined from the separation of the closest detected sources (a few km) up to the field of view ($\sim 100$\,km).
Ionospheric scintillation provides a measurement on the smallest scale, with the approach of \citet{2022JATIS...8a1012R} potentially providing information on intermediate scales.

We note that we have not used the full MWA for our scintillation analysis, but only a compact subarray of 36/128 elements.
This demonstrates that ionospheric scintillation measurements could be made by instruments with much shorter baselines and less collecting area than the full MWA.
For example, SKA-low prototype stations, co-located with the MWA could measure ionospheric scintillation with the significant advantage of all-sky coverage \citep{2021PASA...38...23S}. 

The compact configuration of the MWA Phase II, which retains the core MWA antennas used for our scintillation measurements, but lacks the longer baselines that have been used in previous refractive shift measurements would also be capable of making scintillation measurements.
This motivates us to consider whether ionospheric scintillation measurements can replace refractive shifts as the metric of activity where the latter is unavailable, or difficult to measure.
If we choose a scintillation index of 2\% to set our active vs. quiet threshold, we find that 1043/1447 (72\%) Type I (`quiet') observations are below threshold, while 67\%, 48\%, 94\% of observations are \emph{above} threshold for Types II, III and IV respectively.
Thus, ionospheric scintillation is a reasonable proxy for refractive shifts for assessing ionospheric activity, with only 21\% of `active' observations in total having a `false' negative classification for ionospheric activity based on their scintillation.
Whether the Type I observations with scintillation indices $>$2\% should be considered false positives for ionospheric activity or not depends on whether the structure on the Fresnel scale is a problem in itself.
\citet{chege} found tentative evidence that observations featuring Type I or III ionospheric activity have similar EoR power spectra; this suggests that the anisotropy metric is less important for the purposes of identifying poor, ionospherically contaminated observations. 
We note that the majority of the Type II false negatives have very large refractive shifts, so these might be identified even in data where refractive shifts cannot be measured as easily \citep[such as MWA Phase II compact data; see][]{trott2020}. 
In this case, a combination of all 3 metrics may provide the best indication of ionospheric activity.

Scintillation in the strong regime is more information-rich than weak scintillation due to its spectral structure.
\citet{2020JSWSC..10...10F} exploited this by applying pulsar scintillometry techniques \citep[e.g.][]{2001ApJ...549L..97S} to ionospheric scintillation data to reveal the height and velocity of two distinct scattering regions.

While the MWA cannot access the lower frequencies where strong ionospheric scintillation is common, the exceptionally large number of independent elements on spacings from $\sim10$\,m--$\sim6$,km may provide an alternative approach to deriving more detailed information from weak scattering.
The layout of the MWA Phase II compact configuration, with 3 separate clusters of array elements, each $\sim$100\,m in diameter raises the possibility of using each cluster  to perform a ``multi-station'' analysis.
This is a standard technique in IPS, pioneered by \citet{1967Natur.213..343D}, which was also used to confirm that scintillation due to the interstellar medium was responsible for the intraday variability phenomenon \citep{2000aprs.conf..147J}.
It involves measuring the scintillation signature at different geographical locations. 
These signals are then cross-correlated, and any delay between them can be attributed to the time it takes for the scintillation pattern to drift from one location to another.
With the MWA in compact configuration, the three clusters would be treated as independent sub-arrays.
Any co-located instruments capable of measuring ionospheric scintillation, such as SKA-low prototype stations, could also be used without any requirement that they be \emph{interferometrically} correlated with the MWA.
As well as the tiles that make up the MWA Phase II compact configuration, a Radio Array of Portable Interferometric Detectors \citep[RAPID][]{rapid} or the SKA-low in a mode where each station is treated as a separate interferometer could be used in this way to measure scintillation pattern on the ground. Since this would be an independent and direct measurement of the Fresnel scale, it could be combined with the amplitude scintillation index to determine the height of the irregularities. 

An alternative approach, which would be applicable to both configurations of the MWA, as well as the future SKA-low, would be to measure scintillation via its effect on the complex visibilities.
The theory, as developed by \citet{1972ApJ...174..181C}, shows that for an interferometer which fully samples the scintillation pattern on the ground (i.e. baselines cover a range of values around the Fresnel scale at a range of orientations) the complex visibilities encode not only the strength of scattering but the velocity vector, and even the form of the structure function from the Fresnel scale up to the length of the longest baseline.
Such an approach would, however, likely require observation of a source that is sufficiently bright as to dominate the visibilities.

\subsection{CONCLUSIONS}
\begin{itemize}
    \item We have used ionospheric scintillation in the weak regime to measure the structure of the ionosphere on scales $r_\text{F}\sim$ 300\,m. In the most extreme case observed, the phase variance on this scale was 0.06 rad$^2$ at our observing frequency of 154\,MHz.
    \item The scintillation index is highly correlated (Pearson correlation coefficient 0.71) with refractive shifts measured on a baseline $\sim2$\,km. The relationship between these two metrics suggests that Kolmogorov turbulence is the dominant structure in the ionosphere on scales 300\,m--2\,000\,m, at least in more active ionospheric conditions.
    \item There are a number of observations for which non-turbulent, large-scale structures appear to dominate, but these do not necessarily show anisotropy (and high scintillation indices are also seen when the large-scale structure is highly isotropic)
    \item In the most extreme ionospheric conditions, refractive shifts underestimate the small-scale variance by a factor of four or more.
    \item The more extreme conditions seen in our data could have significant, if manageable, effects on radio astronomy observations. There is much scope and incentive for future work, both to increase our understanding of the mid-latitude ionosphere, and the effects that it may have on trans-ionospheric observations.
\end{itemize}

\section*{Acknowledgements}
We would like to thank C.~Trott for useful discussions regarding the comparison between scintillation and refractive shift measurements.
A.~W acknowledges the invaluable support of an ICRAR summer studentship.
This scientific work makes use of the Murchison Radio-astronomy Observatory, operated by CSIRO.
We acknowledge the Wajarri Yamatji people as the traditional owners of the Observatory site.
Support for the operation of the MWA is provided by the Australian Government (NCRIS), under a contract to Curtin University administered by Astronomy Australia Limited.
We acknowledge the Pawsey Supercomputing Centre which is supported by the Western Australian and Australian Governments.

\begin{table}

\caption{\label{tab:ips}IPS Normalised Scintillation indices of G4JY sources within 15 degrees of the centre of the EoR-0 field. `Fp162' is the GLEAM \citep{gleam2017} flux density at the IPS observing frequency. NSI is the normalised scintillation index as defined by \citet{2018MNRAS.474.4937C} . `Kaplan' indicates the three sources detected by \citet{2015ApJ...809L..12K}.}
\renewcommand{\arraystretch}{0.85} 
\begin{tabular}{|l|l|r|r|l|}
\hline
\multicolumn{1}{|l|}{G4Jy name} &
\multicolumn{1}{c|}{GLEAM name} &
\multicolumn{1}{c|}{Fp162} &
\multicolumn{1}{c|}{NSI} &
\multicolumn{1}{c|}{Kaplan} \\
\hline
G4Jy 2 & GLEAM J000105-165921 & 4.62 & 0.18 & --\\
G4Jy 3 & GLEAM J000312-355630 & 5.02 & 0.35 & --\\
G4Jy 4 & GLEAM J000322-172708 & 10.31 & 0.40 & --\\
G4Jy 6 & GLEAM J000355-305949 & 4.47 & 0.74 & --\\
G4Jy 21 & GLEAM J001218-332157 & 6.23 & 0.38 & --\\
G4Jy 26 & GLEAM J001524-380439 & 9.63 & 0.42 & --\\
G4Jy 28 & GLEAM J001619-143009 & 6.70 & -- & --\\
G4Jy 29 & GLEAM J001636-382643 & 2.92 & -- & --\\
G4Jy 30 & GLEAM J001707-125625 & 5.43 & 0.77 & --\\
G4Jy 33 & GLEAM J001851-124235 & 11.44 & 0.44 & --\\
G4Jy 40 & GLEAM J002112-191041 & 4.71 & 0.20 & --\\
G4Jy 43 & GLEAM J002308-250232 & 8.98 & 0.18 & --\\
G4Jy 45 & GLEAM J002430-292847 & 15.27 & 0.13 & --\\
G4Jy 47 & GLEAM J002530-330336 & 6.27 & 0.87 & --\\
G4Jy 48 & GLEAM J002549-260211 & 18.79 & 1.08 & --\\
G4Jy 49 & GLEAM J002609-124749 & 4.02 & 0.74 & --\\
G4Jy 50 & GLEAM J002613-200455 & 5.73 & 0.21 & --\\
G4Jy 51 & GLEAM J002654-365535 & 4.21 & 0.71 & --\\
G4Jy 64 & GLEAM J003508-200354 & 12.49 & 0.69 & --\\
G4Jy 66 & GLEAM J003629-372745 & 4.21 & 0.35 & --\\
G4Jy 80 & GLEAM J004354-160447 & 4.12 & 0.44 & --\\
G4Jy 81 & GLEAM J004411-221219 & 3.94 & 0.35 & --\\
G4Jy 82 & GLEAM J004441-353029 & 8.49 & 0.51 & --\\
G4Jy 86 & GLEAM J004733-251710 & 6.59 & 0.37 & --\\
G4Jy 106 & GLEAM J005827-240101 & 3.67 & -- & --\\
G4Jy 113 & GLEAM J010241-215227 & 9.19 & -- & --\\
G4Jy 114 & GLEAM J010244-273124 & 5.94 & 0.27 & --\\
G4Jy 1805 & GLEAM J230455-343129 & 4.17 & 0.32 & --\\
G4Jy 1807 & GLEAM J230627-250653 & 5.70 & 0.21 & --\\
G4Jy 1810 & GLEAM J231007-275752 & 6.72 & 0.09 & --\\
G4Jy 1815 & GLEAM J231635-162543 & 6.00 & 0.41 & --\\
G4Jy 1818 & GLEAM J231949-220416 & 5.38 & 0.16 & --\\
G4Jy 1819 & GLEAM J231956-272735 & 8.69 & 0.15 & --\\
G4Jy 1820 & GLEAM J232036-335325 & 3.71 & 0.22 & --\\
G4Jy 1821 & GLEAM J232049-191917 & 4.13 & 0.87 & --\\
G4Jy 1822 & GLEAM J232102-162302 & 16.40 & 1.24 & --\\
G4Jy 1823 & GLEAM J232103-241043 & 4.59 & 0.35 & --\\
G4Jy 1825 & GLEAM J232447-271916 & 4.73 & 1.00 & Y\\
G4Jy 1833 & GLEAM J232831-145415 & 6.70 & 0.23 & --\\
G4Jy 1834 & GLEAM J232834-210542 & 4.98 & 0.08 & --\\
G4Jy 1835 & GLEAM J232933-192257 & 6.76 & 0.09 & --\\
G4Jy 1836 & GLEAM J233003-180803 & 3.89 & 0.28 & --\\
G4Jy 1837 & GLEAM J233147-245210 & 5.85 & 0.17 & --\\
G4Jy 1839 & GLEAM J233343-305753 & 4.92 & 1.05 & Y\\
G4Jy 1841 & GLEAM J233429-334502 & 3.79 & 0.15 & --\\
G4Jy 1844 & GLEAM J233648-344403 & 8.00 & 0.13 & --\\
G4Jy 1846 & GLEAM J234112-162052 & 9.26 & 0.31 & --\\
G4Jy 1848 & GLEAM J234146-350620 & 9.43 & 0.86 & Y\\
G4Jy 1849 & GLEAM J234324-214129 & 4.52 & 0.08 & --\\
G4Jy 1850 & GLEAM J234545-240232 & 3.66 & 0.12 & --\\
G4Jy 1852 & GLEAM J234740-280839 & 3.69 & 0.31 & --\\
G4Jy 1854 & GLEAM J235050-245702 & 9.41 & 0.15 & --\\
G4Jy 1855 & GLEAM J235134-160739 & 5.24 & 0.32 & --\\
G4Jy 1858 & GLEAM J235701-344532 & 19.65 & 0.03 & --\\
G4Jy 1860 & GLEAM J235708-181743 & 4.03&  & --\\
\hline\end{tabular}
\end{table}

\bibliography{refs}

\end{document}